# Superconductivity in the vicinity of antiferromagnetic order in CrAs

Wei Wu[1†], Jinguang Cheng[1,2†*], Kazuyuki Matsubayashi[2], Panpan Kong[1], Fukun Lin[1], Changqing Jin[1], Nanlin Wang[1], Yoshiya Uwatoko[2*], and Jianlin Luo[1*]

[1] *Beijing National Laboratory for Condensed Matter Physics and Institute of Physics, Chinese Academy of Sciences, Beijing 100190, China*

[2] *Institute for Solid State Physics, University of Tokyo, 5-1-5 Kashiwanoha, Kashiwa, Chiba 277-8581, Japan*

[†]These authors contributed equally to this work.

*E-mail: jgcheng@iphy.ac.cn; uwatoko@issp.u-tokyo.ac.jp; jlluo@iphy.ac.cn

**One of the common features of unconventional, magnetically mediated superconductivity as found in the heavy-fermions, high-transition-temperature (high-$T_c$) cuprates, and iron pnictides superconductors is that the superconductivity emerges in the vicinity of long-range antiferromagnetically ordered state.[1] In addition to doping charge carriers, the application of external physical pressure has been taken as an effective and clean approach to induce the unconventional superconductivity near a magnetic quantum critical point (QCP).[2,3] Superconductivity has been observed in a majority of 3d transition-metal compounds,[4-9] except for the Cr- and Mn-based compounds in the sense that the low-lying states near Fermi level are dominated by their 3d electrons. Herein, we report on the discovery of superconductivity on the verge of antiferromagnetic order in CrAs via the application of external high pressure. Bulk superconductivity with $T_c \approx 2$ K emerges at the critical pressure $P_c \approx 8$ kbar, where the first-order antiferromagnetic transition at $T_N \approx 265$ K under ambient pressure is completely suppressed. Abnormal normal-state properties associated with a magnetic QCP have been observed nearby $P_c$. The close proximity of superconductivity to an antiferromagnetic order suggests an unconventional pairing mechanism for the superconducting state of CrAs. The present finding opens a new avenue for searching novel superconductors in the Cr and other transitional-metal based systems.**

CrAs adopts the orthorhombic MnP-type structure (inset of Fig. 1) with unit-cell parameters $a$ = 5.649Å, $b$ = 3.463 Å, and $c$ = 6.2084 Å at room temperature. Its crystallographic and magnetic properties have been extensively investigated during 1970s due to the observation at $T_N$ = 270 ± 10 K of a first-order antiferromagnetic transition,[10-12] which is accompanied with discontinuous changes of lattice parameters. The most dramatic change is a sudden expansion of lattice constant $b$ by ~ 4% below $T_N$.[13] Neutron diffraction measurements established a double-helical magnetic structure propagating along the orthorhombic $c$ axis with the Cr moment of 1.7 $\mu_B$ lying essentially within the $ab$ plane.[11,12] The acquisition of large Cr moment below $T_N$ has been rationalized as an itinerant- to localized-electron transition as a consequence of the sudden elongation of Cr-Cr seperation along the $b$ direction.[10] The coupled structural, magnetic, and electronic degrees of freedom at $T_N$ make high pressure a very effective knob in fine tuning its ground-state properties. Indeed, earlier high-pressure studies have shown that the antiferromagnetic transition at $T_N$ can be readily suppressed by pressure and vanishes completely at P ~ 5 kbar.[14] Since exotic phenomena such as the unconventional superconductivity are frequently observed in the vicinity of a magnetic quantum critical point, we are motivated to explore in great detail the physical properties of antiferromagnetic CrAs as $T_N$ is suppressed progressively under hydrostatic pressure.

We measured the resistivity ρ(T) of a number of CrAs single crystals under various hydrostatic pressures up to 21.4 kbar in a wide temperature range from 300 K down to 70 mK. The ρ(T) data of one represented sample shown in Fig. 1(a) illustrated the variation of $T_N$ with pressure (See the **Method** section about how the pressure values at $T_N$ are estimated). As seen in Fig. 1(a), the first-order antiferromagnetic transition is manifested as a sudden drop of resistivity at $T_N$ = 264 K upon cooling at ambient pressure. As indicated by the arrows, $T_N$ decreases progressively with increasing pressure; the resistivity anomaly at $T_N$ changes from a sudden drop for P < 3 kbar to a smoothly-varied kink whose magnitude is also diminished gradually within the pressure range 3 < P < 7 kbar. Such an evolution of the resistivity anomaly at $T_N$ suggests a gradual crossover from strong to weak first-order transition, especially near the critical pressure where $T_N$ vanishes. It should be noted that in the previous high-pressure study on a polycrystalline sample $T_N$ terminated at about 150 K under 5 kbar,[14] whereas our measurements on a high-quality single-crystalline sample enabled us to track $T_N$ down to 70 K at P = 6.97 kbar. Above this pressure, no anomaly can be discerned in resistivity above 3 K.

Our key finding of this study is the observation of superconductivity in CrAs with the application of high pressure. The $\rho(T)$ data shown in Fig. 1(b, c) depict explicitly how the superconductivity emerges and then evolves with pressure at low temperatures. Below $T_N$, $\rho(T)$ measured outside the pressure cell does not show any anomaly down to 350 mK, thus confirming the absence of superconductivity in CrAs at ambient pressure. Interestingly, when the sample was subjected to a minute pressure below 3 kbar, $\rho(T)$ starts to decrease below ~2.5 K and displays multi-step drops before reaching a constant, non-zero value near 1 K. The amplitude of resistivity drop grows progressively with pressure, and eventually reaches zero resistivity near 1 K at P ≈ 3 kbar, signaling the occurrence of superconductivity. Upon further increasing pressure in the range 3 < P < 7 kbar where $T_N$ remains finite, the superconducting transition is featured by similar multi-step drops; the onset temperature first increases and then decrease with pressure, while the zero-resistivity temperature $T_c^0$ ≈1 K keeps nearly constant. The multi-step feature fades away with further increasing pressure, and is finally changed to a single transition for P > 8 kbar where $T_N$ vanishes completely. As seen in Fig. 1(c), $T_c^0$ first increases and reaches a maximum of 1.47 K at P = 10.88 kbar, above which it decreases gradually and reaches 1.17 K at 21.37 kbar, the maximum pressure of the present study. It should be noted that the pressure-induced superconductivity in CrAs as well as the peculiar features of resistivity shown in Fig. 1 are well reproducible as confirmed on three independent CrAs single crystals of similar quality, *i.e.* the residual resistivity ratio RRR > 200.

The above resistivity results revealed that superconducting state coexists with the antiferromagnetically ordered state within a broad pressure range below 7 kbar; the progressive development of the resistivity drop and the multi-step feature of superconductivity transition signal the growth of superconducting state at the expense of the antiferromagnetic state. In order to gain insights into the competing nature of these two states and to further verify the bulk nature of the observed superconductivity, we resorted to ac magnetic susceptibility measurements that can probe the diamagnetic signal due to superconducting transition. As shown in Fig. 2, the diamagnetic signal starts to appear at P > 3 kbar, and the superconducting shielding fraction grows steadily with pressure, reaching over 90% of the sample volume above 8 kbar. These results thus confirmed directly the bulk nature of pressure-induced superconductivity in CrAs. In addition, the onset temperatures of the diamagnetic signal are also in excellent agreement with the $\rho(T)$ data shown in Fig. 1(c).

Based on the above results, we can construct a temperature-pressure (T-P) phase diagram for CrAs single crystal. As seen in Fig. 3, the first-order antiferromagnetic transition temperature $T_N$ = 264 K at ambient pressure can be suppressed quickly by the application of external pressure; an extrapolation of $T_N(P)$ curve gives a critical pressure $P_c \approx$ 8 kbar where $T_N$ approaches zero temperature. At $P \geq P_c$, bulk superconductivity with shielding fraction more than 90% is realized accompanied with an order of magnitude decrease of the 10-90% superconducting transition temperature width $\Delta T_c$, as seen in Fig. 3(b). In this pressure region, the superconducting transition temperature $T_c$ exhibits a broad maximum around 11 kbar and then decreases gradually with further increasing pressure. For 3 kbar < P < $P_c$, the superconducting state coexists with the antiferromagnetic state, and the volume fraction grows gradually at the expense of the other. It is interesting to note that the superconducting state in this pressure region is characterized by a higher onset temperature for superconductivity, Fig. 1(b).

The phase diagram of CrAs shown in Fig. 3 resembles that of many unconventional superconducting systems,[2,15,16] including the heavy-fermion, high-$T_c$ cuprate, and iron-pnictide superconductors. As a universal trend in these systems, superconductivity emerges in the vicinity of a quantum critical point (QCP) where a high-temperature ordered state involving spin, charge, or lattice degrees of freedom is suppressed via applying external tuning parameter $\delta$, such as doping charge carrier, chemical or physical pressure. In addition, the superconducting transition temperature $T_c$ always passes through a maximum value at some critical $\delta_c$, leading to a dome-shaped $T_c$-$\delta$ phase diagram. The proximity of superconductivity to a QCP lends strong support for the unconventional mechanism for Cooper pairing via the critical fluctuations. Despite the first-order nature of the antiferromagnetic transition in CrAs, the following observations suggested the presence of strong antiferromagnetic fluctuations near the critical pressure $P_c$ and the possible unconventional pairing mechanism.

First, we found that within a large pressure regime 3 < P < 21 kbar where superconductivity takes place the normal-state $\rho(T)$ below 10 K follows the power-law relationship: $\rho(T) = \rho_0 + AT^n$ with the exponent *n* falling in the range 1.5 ± 0.1, which is very close to the value predicated for incoherent scattering of quasiparticle via magnetic interactions in three-dimensional antiferromagnets.[17,18] One of the representative $\rho(T)$ data at 9.5 kbar is plotted in the form of $\rho$ vs $T^{1.5}$ in Fig. 3(c). In our previous study,[19] we have shown that $\rho(T)$ of CrAs single crystal at ambient pressure follows the Fermi-liquid behavior below 10 K, with the Kadowaki-Woods ratio $A/\gamma^2$ falling onto the universal line of many correlated metals, where $\gamma$ is the electronic specific-heat coefficient. Our fitting to the ambient $\rho(T)$ confirms the previous results with n = 2.15(4). The observation of n < 2 has been taken as a characteristic signature for non-Fermi-liquid metals near a magnetic QCP.[20]

Second, the antiferromagnetic fluctuations seem to be abundant above $T_N$ in that the magnetic susceptibility of CrAs keeps increasing with temperature up to at least 700 K under ambient pressure.[19] As a matter of fact, such an increasing behavior of magnetic susceptibility has been found to be universal in the recently discovered iron-based superconductors, and regarded as an indication for antiferromagnetic fluctuations.[16] In this sense, these spin fluctuations would be optimized near the critical pressure where the antiferromagnetic order of CrAs is suppressed completely. Therefore, the critical spin fluctuations associated with the quantum criticality could act as an important glue medium for Cooper pairing.

Last but not least, the pressure-induced superconductivity in CrAs is found to be sensitive to the residual resistivity $\rho_0$. Our initial high-pressure studies on the CrAs crystals with $\rho_0 \approx 10$ μΩ cm (RRR ≈ 40-50) displayed indications of superconductivity below 1 K, but no zero resistivity can be achieved. We finally confirmed the bulk superconductivity with higher $T_c$ on high-quality CrAs crystals having much reduced $\rho_0 \approx$ 1-2 μΩ cm. Such a sensitivity of superconductivity and the associated $T_c$ to disorders (impurities and/or defects) would imply an unconventional (non s-wave) paring mechanism as found in heavy-fermion superconductors[2] and the p-wave superconductor $Sr_2RuO_4$.[21] In these cases, superconductivity is expected to be destroyed when the electron mean free path $l_{mfp}$ is smaller than the superconducting coherent length $\xi$. Our rough estimations of $l_{mfp}$ and $\xi$ seem to verify such a scenario. We first estimated $\xi$ near the optimal pressure from the upper critical field $\mu_0 H_{c2}$, which is determined from the $\rho(T)$ curves under different magnetic fields. Fig. 4 displayed the $\mu_0 H_{c2}$ as a function of $T_c$ at P = 9.5 kbar. Here, we determined $T_c$ as the middle point between 10% and 90% drop of resistivity with the error bar being the 10-90% width. As shown in Fig. 4, $\mu_0 H_{c2}$ versus $T_c$ can be fitted well with the empirical formula: $\mu_0 H_{c2}(T) = \mu_0 H_{c2}(0) \{1-[T_c/T_c(0)]^\alpha\}$, with $\mu_0 H_{c2}(0) = 0.96(2)$ T, $T_c(0) = 1.506(6)$ K, and $\alpha = 1.41(6)$, respectively. The obtained $\mu_0 H_{c2}(0)$ allows us to estimate the Ginzburg-Landau coherence length $\xi = 185$ Å according to the relationship: $\mu_0 H_{c2}(0) = \Phi_0/2\pi\xi^2$, where $\Phi_0 = 2.067 \times 10^{-15}$ Web is the magnetic flux quantum. Based on the measured Hall coefficient $|R_H| = 2.6 \times 10^{-10}$ m$^3$/C at 5 K, on the other hand, $l_{mfp}$ values of about 766 Å and 153 Å are obtained for the clean ($\rho_0 = 2$ μΩ cm) and dirty ($\rho_0 = 10$ μΩ cm) samples, respectively, by assuming a spherical Fermi surface with one kind of charge carrier. In spite of large uncertainty in the above estimations, we confirmed that bulk superconductivity can be seen only in samples with $l_{mfp} > \xi$, a signature for unconventional pairing mechanism.

Although superconductivity has been reported for some chromium alloys,[22] *e.g.* Cr-Rh and Cr-Re, to the best of our knowledge, *CrAs is the first superconductor among the Cr-based compounds*. Unlike the alloys, most of the density of states near Fermi level for CrAs is attributed to the Cr-3d states,[23] which makes the magnetic degree of freedom relevant for the observed superconductivity. As a matter of fact, the itinerant-electron antiferromagnetism of chromium metal itself has been the subject of much interest over fifty years.[24] To date, no unambiguous evidence of superconductivity has been observed near the quantum critical point when the spin-density-wave antiferromagnetism of Cr is suppressed by either high pressure[25,26] or chemical doping[27]. In this regard, further theoretical and experimental studies are needed to elucidate the pairing symmetry and the dominant mechanism, especially the role of magnetism, for the observed superconductivity in CrAs.

CrAs belongs to the large family of transition-metal pnictides with a general formula MX (M = transition metal, X = P, As, Sb), which form in either hexagonal NiAs-type (B8$_1$) or orthorhombic MnP-type (B31) structure (for a review, see Ref. 28 and references therein). Previous studies have demonstrated a close relationship between the magnetism of 3d-element and the M-M interatomic distance due to the direct orbital overlap across the shared MX$_6$ edges. With decreasing the size of X atom in the CrX series, the Cr moment decreases from the nearly full moment of 3 μ$_B$/Cr for X = Sb to 1.7 μ$_B$/Cr for X = As, and finally to nonmagnetic for X = P.[28] It seems that the magnetism of Cr in CrAs is located very close to the boundary of localized- to itinerant-electron transition because a 5%-P doping has been shown to suppress completely the long-range antiferromagnetic transition. This fact can explain why the antiferromagnetic transition is so sensitive to the external pressure as we observed in this study. Beside, chemical substitutions for Cr in $Cr_{1-x}M_xAs$ can also suppress effectively the antiferromagnetic transition. Therefore, our discovery of superconductivity in CrAs single crystal would stimulate revival interest in this family of compounds in the context of possible unconventional superconductivity.

## Methods

CrAs single crystals were grown out of the Sn flux. Details about the crystal growth procedures and sample characterizations at ambient pressure have been given elsewhere.[19] High pressures were generated with a self-clamped piston-cylinder cell made of non-magnetic BeCu and NiCrAl alloys.[29] Glycerol with a room-temperature solidification pressure far above 2 GPa was chosen as the pressure transmitting medium in order to minimize the shear stress when changing pressures at room temperature. The pressure (P) inside the piston-cylinder cell was monitored by detecting the superconducting transition temperature ($T_c$) of a piece of Pb according to the equation: P (kbar) = ($T_0$ - $T_c$)/0.0365, where $T_0$ = 7.20 K is the $T_c$ of Pb at ambient pressure. It should be noted that the pressure inside the clamp-type pressure cell is not constant within the whole temperature range. It has been shown that the pressure drops about 3 kbar upon cooling down from room temperature to 4.2 K.[30] Since the antiferromagnetic transition of CrAs spans over a large temperature range under pressures, we have to estimate the pressure at different temperatures properly. According to Ref. 30, the pressure is nearly constant below 90 K and it increases linearly to 300 K with a slope of 0.012 kbar/K for pressure below 8 kbar. The pressure values at $T_N$ shown in Fig. 1(a) are estimated by the extrapolation: P($T_N$) = P(Pb) + 0.012 ($T_N$-90).

The low temperatures down to 70 mK were reached by attaching the pressure cell to a Heliox-AC insert (Oxford Instrument) or dilution refrigerator. The resistivity was measured by the conventional four-probe method with the current applied along the orthorhombic *b* axis. Ac magnetic susceptibility was measured with a mutual induction method at a fixed frequency of 317 Hz with a modulation field of about 2 Oe. The diamagnetic signal due to the superconductivity transition was estimated by comparing to the diamagnetic signal of Pb, which served as a pressure manometer, with nearly the same size as the CrAs sample.

## Acknowledgements

We thank Y. P. Wang, Z. Fang, J. Q. Yan, J. S. Zhou, J. B. Goodenough, N. Mori, T. Kato, T. Kanomata, M. Matsuda, F. Steglich, Q. Si, P. J. Sun, G. M. Zhang and L. Yu for fruitful discussions. Work at IOP/CAS was supported by the National Science Foundation of China (Grant Nos. 11025422, 11304371), the National Basic Research Program of China (Grant Nos. 2014CB921500, 2011CB921700), and the Strategic Priority Research Program of the Chinese Academy of Sciences（Grant Nos. XDB01020300, XDB07000000). Work at ISSP/UT was partially supported by Grant-in-Aid for Scientific Research, KAKENHI (Grant Nos. 23340101, 252460135), the Grant-in-Aid for Young Scientist B (No. 24740220) from the Ministry of Education, Culture, Sports, Science and Technology, and the JSPS fellowship for foreign researchers (Grant No. 12F02023).

## Author contributions

W.W., J.L. and N.W. synthesized the high-quality CrAs single crystals and initiated the present study; J.C., and K.M. measured the resistivity and ac magnetic susceptibility under high pressure with piston-cylinder cell and confirmed the bulk superconductivity; F.L. performed resistivity, dc magnetic susceptibility, and Hall coefficient measurements at ambient pressure; P.K. and C.J. carried out the initial high-pressure resistivity measurements with diamond anvil cell; J.C. and J.L. wrote the paper with contributions from all authors. J.L. and Y.U. supervised the project.

## Competing financial interests

The authors declare no competing financial interests.

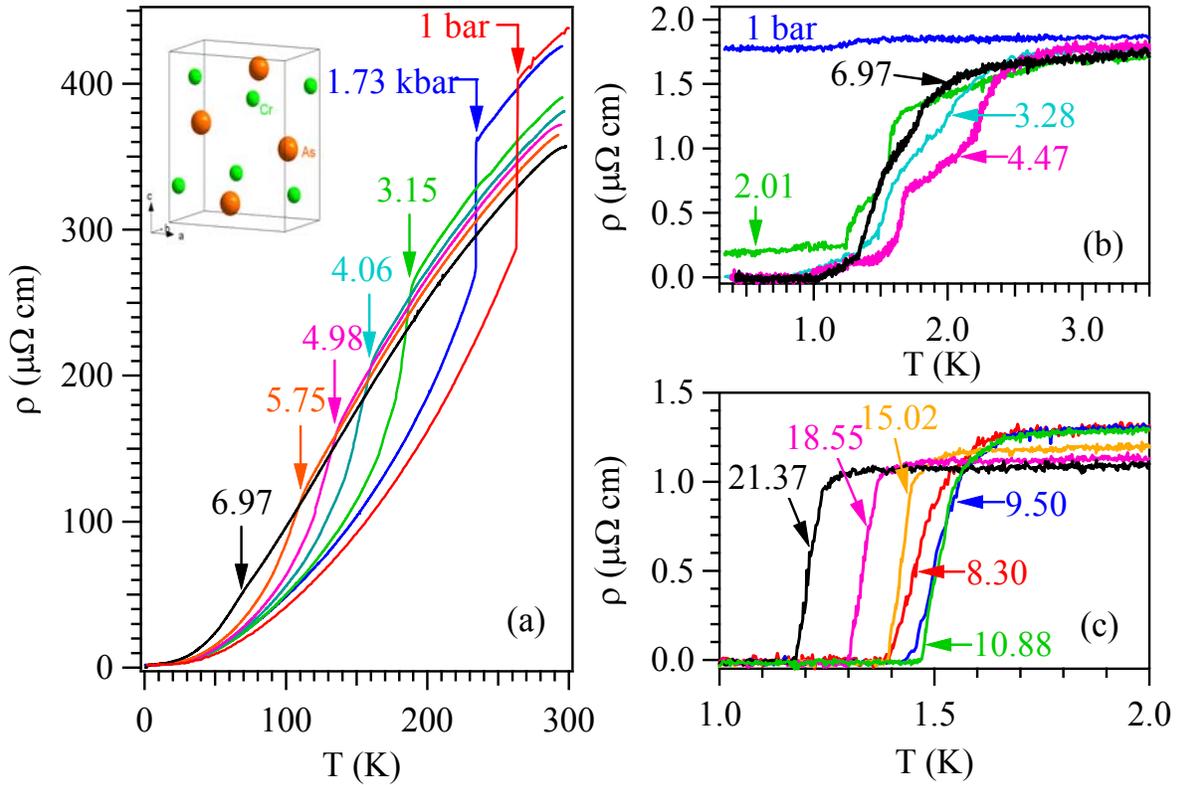

**Figure 1** (color online) Resistivity ρ(T) of the CrAs single crystal under various hydrostatic pressures. (a) ρ(T) data up to P = 6.97 kbar in the whole temperature range highlighting the variation with pressure of the antiferromagnetic transition temperature $T_N$ indicated by the arrows, (b) and (c) ρ(T) below T = 4 K in the whole investigated pressure range highlighting the evolution with pressure of the superconducting transition. The pressure values listed in (a) are the estimated pressure at $T_N$, see **Method** for details. The pressure values given in (b) and (c) are determined directly from the superconducting transition of Pb. A schematic crystal structure of MnP-type CrAs is displayed in the inset of Fig. 1(a).

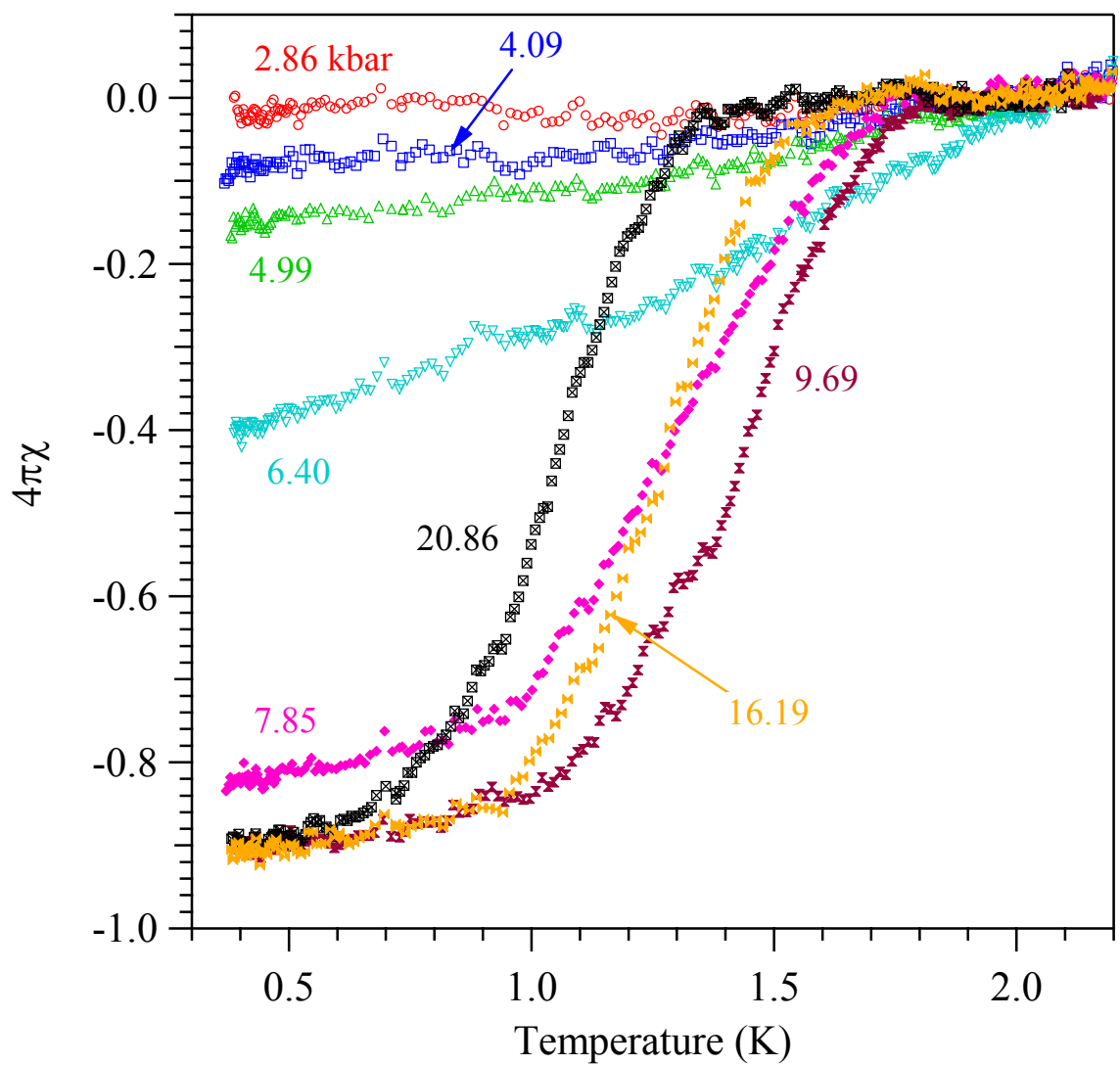

**Figure 2** (color online) Temperature dependence of the ac magnetic susceptibility $4\pi\chi$ of a CrAs single crystal under various hydrostatic pressures.

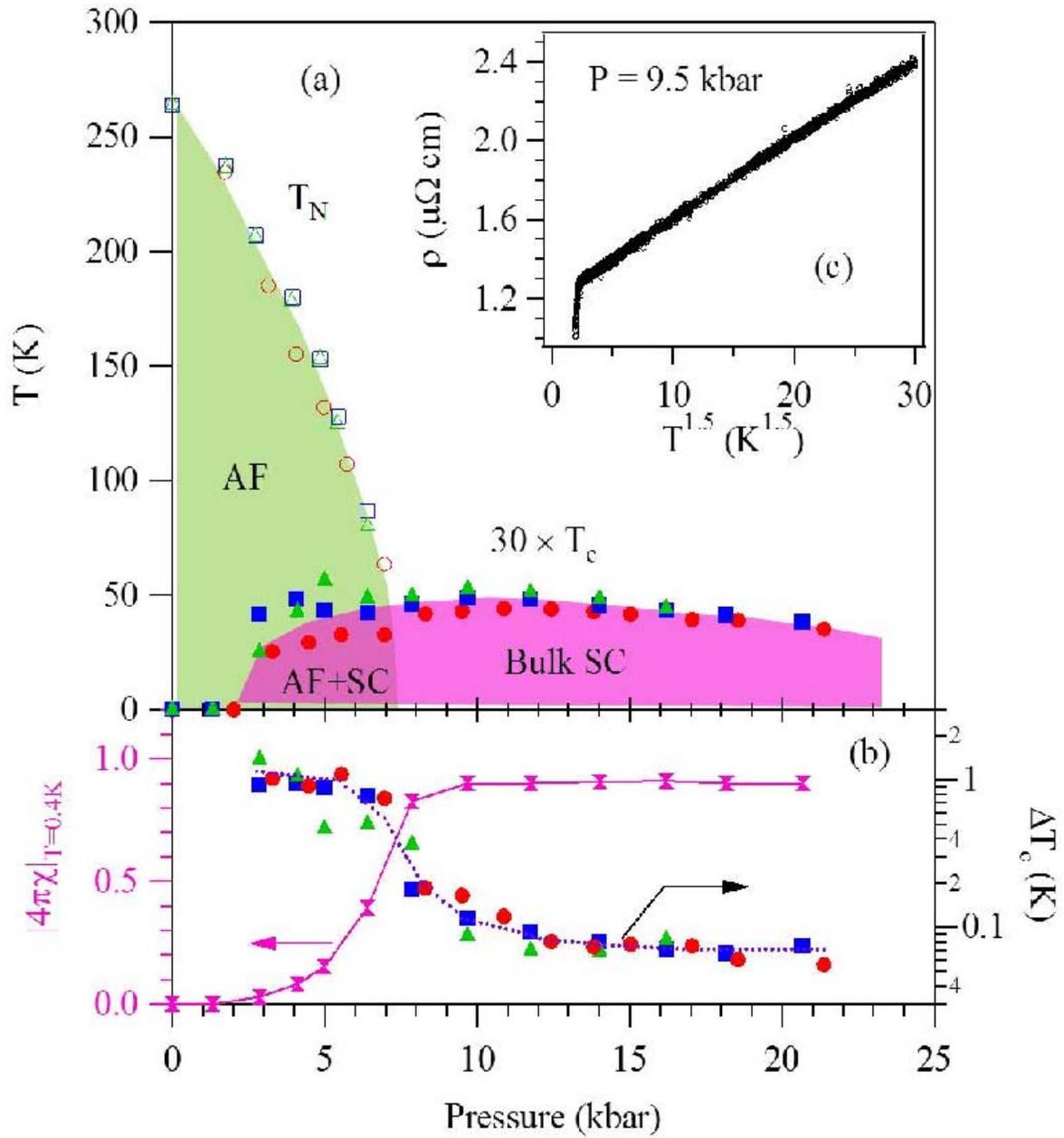

**Figure 3** (a) Temperature-pressure phase diagram of CrAs; (b) the superconducting shielding fraction and transition temperature width as a function of pressure; the symbols of circle (red), square (blue), and triangle (green) in (a, b) represent three independent samples with RRR = 240, 327, and 250, respectively. (c) A plot of $\rho$ vs $T^{1.5}$ for the normal-state $\rho(T)$ curve at P = 9.5 kbar.

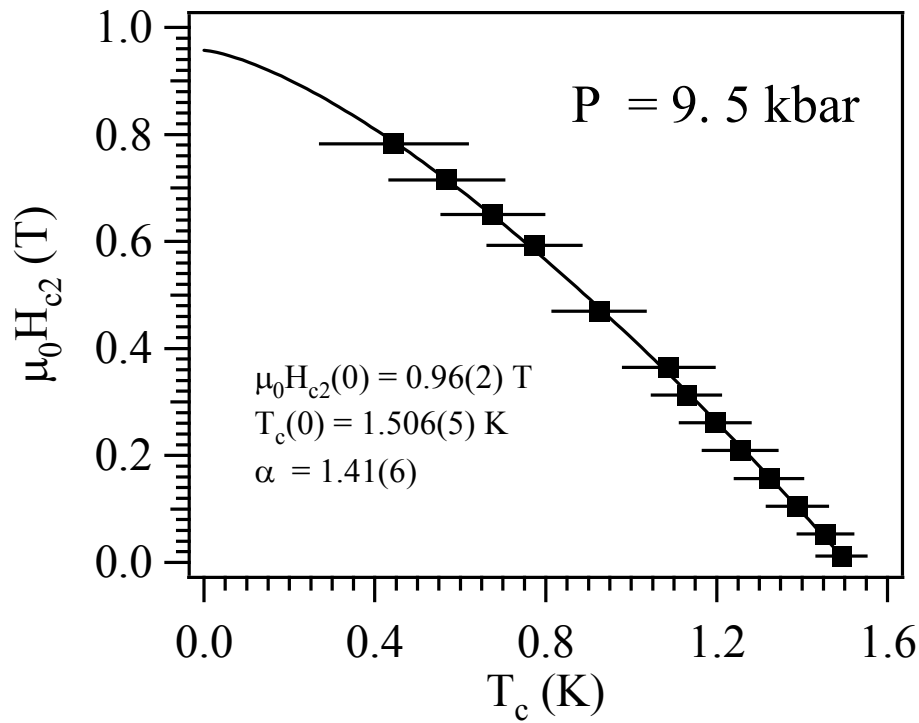

**Figure 4** Temperature dependence of the upper critical field $\mu_0H_{c2}$ for CrAs single crystal at P = 9.5 kbar. Solid line is the fitting curve, see the text for details.